\documentstyle[12pt]{article}
\begin{document}
\title{Alfven Waves in Disks, Outflows and Jets}
\author{Reuven Opher\thanks{email: opher@astro.iag.usp.br}\\
Instituto de Astronomia, Geof\'\i sica e Ci\^encias Atmosf\'ericas\\
Universidade de S\~ao Paulo\\
Rua do Mat\~ao 1226, Cidade Universit\'aria\\
CEP 05508-900 S\~ao Paulo, S.P., Brazil}
\maketitle
\begin{abstract}
Observations of young stellar objects, such as classical T Tauri stars (CTTSs),
show evidence of accretion disks (ADs), outflows and jets. It is shown that Alfven 
waves (AWs) are important in all of these phenomena. AWs can be created by turbulence 
in the AD and magnetosphere of a CTTS.
Jets, stellar outflows, and the heating of the central regions of ADs may
be due to the damping of AWs. Our recent investigations of the heating of the 
AD and the magnetosphere by AWs as well as of the structures produced in outflows 
from ADs initiated by turbulent AWs, are discussed. We comment on the possible 
importance of AWs in resolving the energy and angular momentum problems of CTTSs.
\end{abstract}
\par
\section{Introduction}
The current picture of a classical T Tauri star (CTTS) is that of a central protostar, 
surrounded by a thin accretion disk that was formed as the result of the collapse of 
matter in a molecular cloud. At the corotation radius, the disk is disrupted due to 
the star's dipole magnetic field.  The accreting matter then follows the star's magnetic 
field lines until it impacts on the stellar surface. This magnetospheric model accounts 
for the observational signatures seen in CTTSs, such as the excess of optical and 
ultraviolet continuum flux (veiling) and redshift absorption features in the 
emission line profiles (inverse P Cygni profiles) (Muzerolle, Hartmann, \& 
Calvet 1998; Hartmann, Hewitt, \& Calvet 1994). CTTSs display both outflow and 
inflow signatures (see, e.g., Edwards, Ray, \& Mundt 1993). 
\par
Magnetic field lines on the order of 1 kG at the surface of the protostar are sufficient
to disrupt the disk. The existence of magnetic fields of this magnitude on the star's 
surface is inferred from observations (Johns-Krull et al. 1999; Guenther et al. 1999). 
Coupling of the accreting matter to the star's magnetic field lines is possible
if the temperature of the disk at the truncation radius is $>10^3\;K,$ at which collisional 
ionization of metal ions comes into effect (Umebayashi \& Nakano 1988).
\par
A major problem to be resolved concerns the gravitational energy $E_G$ of the 
accreted matter, which is over an order of magnitude greater than the emitted energy 
$E_{\ast}=\bar{L}_\ast\tau_{\ast}$ of the CTTS, where $\bar{L}_{\ast}$ is the average optical 
luminosity and $\tau_{\ast}$ is its lifetime. It has been suggested that 
the occurance of large ejections could possibly get rid of the excess gravitational energy, 
but no detailed analysis has been made so far. 
\par
Another problem that must be dealt with is the large amount of angular momentum loss 
required for the formation of a protostar in a CTTS. The average protostellar mass 
$M_{\ast}$ in a molecular cloud has an angular momentum 4-5 orders of magnitude 
greater than a protostar in a CTTS. It has been suggested that the Balbus and Hawley 
Instability (BHI) (Balbus \& Hawley 1991, 1998) in the accretion disk may be able to 
get rid of the excess angular momentum. Detailed calculations of this possibility have 
yet to be made, although it has been shown in 3D numerical simulations that angular 
momentum can, in fact, be transported outward in the disk by the BHI (Hawley \& Stone 1998).
\par
Energy and angular momentum in CTTSs can be transported over large distances by Alfven 
waves which are generated by perturbations of the magnetic field, embedded in the plasma 
of the protostar. In general, AWs are weakly damped, as compared with sonic waves.
The wave transports the energy and angular momentum away from the source of the 
perturbation, generally turbulence, along the magnetic field lines at the Alfven velocity. 
Classical plasma processes do not allow for accretion to take place in a disk. It is, thus, 
generally assumed that turbulence exists in the disk, in order to produce the anamolous 
viscosity required for accretion. This assumption justifies the existence of AWs in the disk.
\par
AWs have indeed been observed in a space physics environment, in particular, 
the solar wind. Near the sun, most of the fluctuations observed are propagating outwards
(Bavassano 1990). Tu, Marsch and Thieme (1989) and Tu, Marsch and Rosenbauer (1990) 
found that, at a distance $\sim 64 R_{\odot}$ from the sun, the fluctuations are 
outward-going with periods of $\sim\rm{minutes}$ to days. All of the observations indicate
that these fluctuations are, in fact, AWs. The waves are observed to have a 
power law distribution. Jatenco-Pereira, Opher and Yamamoto (1994) showed that the observed 
AW flux can explain the observed velocity of the solar wind as well as the
observed radial temperature distribution in the corona and chromosphere. In the calculations 
of Jatenco-Pereiro et al. (1994), the network heating of Parker (1991) with a flux of 
$8.7\times 10^5\;\rm{ergs\, cm^{-2}\, s^{-1}},$ which has a damping length of
$0.5\;R_{\odot}$, was used. An Alfven flux of $1.3\times 10^5\;\rm{ergs\, cm^{-2}\, s^{-1}}$ 
was found to be necessary in order to fit the observed data. A coronal hole geometry, 
in which surface Alfven damping (Lee and Roberts 1986, Hasegawa and Uberoi 1982) at 
the border was assumed, was studied.  The calculations predicted a spectral break at 
$10^{-2}\; \rm{Hz}$ for a distance of $64 R_{\odot},$ as is observed. This Alfven flux 
produced a $\sim 600\;\rm{km s^{-1}}$ wind, as is also observed.
\par
In Section 2, our recent work on AW heating of the accretion disk and 
magnetosphere of CTTSs is presented. The structures produced in outflows from 
accretion disks, initiated by turbulent AWs, are treated in Section 3.
Concluding remarks and a discussion of the energy and angular momentum problems in CTTS
are given in Section 4.
\par
\section{AW heating in the accretion disk and magnetosphere of a CTTS} 
Accretion in the disk about a CTTS is generally assumed to be due to the BHI. An ionized 
plasma is required in order for the BHI to occur. However, Gammie (1996) predicted that 
the accretion disk is unionized in its central region from 0.1 to 5 AU. Based on the
evidence by Stone et al. (1996) that the BHI becomes turbulent, Vasconcelos, 
Jatenco-Pereira and Opher (2000) suggested that the BHI turbulence in the ionized regions 
produce AWs that propagate into the central region and heat it, so that the 
entire disk becomes ionized. They found that, assuming nonlinear mode coupling for 
the damping mechanism of an AW with a frequency 1/10 the ion cyclotron frequency, 
an AW amplitude of only 0.2\% of the Alfven velocity was sufficient to create 
the necessary heating in the disk.
\par
The required temperature profile of the magnetosphere, based on observations, was calculated 
by Hartmann, Hewitt and Calvet (1994). Martin (1996) studied the heating of the accreting 
matter as it follows the magnetic field lines in the magnetosphere. He took into account 
adiabatic compression, photoionization and ambipolar diffusion. He found too low a 
temperature for the magnetosphere due to these heating mechanisms and concluded that an 
additional heating source must be present. Vasconcelos, Jatenco-Pereira and Opher (2002) 
found that for an Alfven frequency of 1/10 the ion cyclotron frequency, an AW
amplitude of only 0.3\% the Alfven velocity was required to attain the necessary temperature.
\par
\section{Structures produced in the turbulent 
AW-initiated outflows from accretion disks}
In a cold, non-turbulent disk, matter cannot leave the surface since it is bound 
gravitationally. A magnetic field perpendicular to the AD is generally assumed in CTTSs. 
Vitorino, Jatenco-Pereira and Opher (2002, 2003) studied the outflow from the accretion disk 
in a CTTS along the magnetic field lines, assuming, as did Ouyed and Pudritz (1997), that 
the corona of the disk is supported by a turbulent AW pressure $P_A$, taken to be 
$0.39 P_{KR},$ where $P_{KR}$ is the Keplerian ram pressure, defined as $\rho V_K^2,$ and 
$\rho$ and $V_K$ are the density in the disk and the Keplerian velocity, respectively. The 
$P_A$ balances the gravitational attractive force of the disk, allowing outflow to take 
place due to the centrifugal force. It is generally assumed that the twisting of the magnetic 
field lines due to the rotation of the disk will eventually transform the outflow into a jet.
\par
Gas pressure in the corona $P_G$ was taken to be $0.01 P_{KR}.$ The initial poloidal magnetic
field, perpendicular to the disk, was assumed to be $(8\pi P_G)^{1/2}.$ It was assumed that 
the magnitude of the initial toroidal field was equal to that of the initial poloidal field, 
to which it was perpendicular. The density in the corona was taken to be $0.01\rho.$
\par
The structure of the outflow after several hundred rotations of the inner Keplerian orbit was 
studied by Vitorino et al. (2002, 2003), using a 3D ZEUS code. Initially, they perturbed the 
outflow with a random velocity amplitude proportional to $r^{-a},$ where $r$ is the radius in 
the accretion disk and $a=-3/2,\; -1,\; -1/2,\; 0,\; 1/2$ and 1. The maximum random velocity
perturbation at the inner orbit of the disk was 0.01 the Keplerian velocity. Although the 
perturbation was random, all the structures that formed had a spacing of 11 times the radius 
of the inner Keplerian orbit.
\par
Periodic perturbations of the outflow were also studied by Vitorino et al. (2003). In this 
case, the structures that formed had a spacing of $T/2$ times the radius of the
inner Keplerian orbit, where $T$ is the period of the perturbation. They investigated periods 
10-80 times that of the inner Keplerian orbit $T_{Ki}$ and found that the structures 
dissipated for small $T (\sim 10-20 T_{Ki})$ and tended to fragment for large
$T (\sim 60-80 T_{Ki}).$
\par
It is to be noted that the investigations of Vitorino et al. (2000, 2003) are relevant for 
the formation of structures very near to the source, as yet unresolved with present telescopes. 
For example, the spacing of 11 times the radius of the inner Keplerian radius corresponds to a 
distance of only $\sim 10^{-6}\;\rm{pc}.$
\par
\section{Conclusions}
In Section 2, it was shown that AWs may be important in ionizing the central unionized 
region of an AD of a CTTS, making it possible for the BHI to operate and accretion to take 
place. It was also shown in Section 2 that AWs may be important in raising the 
temperature of the magnetosphere of a CTTS, so that it agrees with observations. In 
Section 3, turbulent AW-initiated outflow from the disk was seen to create 
interesting structures.
\par
Finally, I would like to comment on the energy and angular momentum problems with respect
to CTTSs which were discussed in the introduction. Concerning the problem of the excess
angular momentum, it is possible that it may be transported away from the disk by AWs 
instead of the BHI. As noted, AWs can transport energy and angular momentum over great 
distances along magnetic field lines. Moreover, turbulent AWs can not only initiate mass 
outflow, as discussed in Section 3, but can also carry angular momentum out of the disk, 
depositing it in the molecular cloud. Concerning the energy problem, the major portion of 
the gravitational energy of the accreted matter is lost when its distance to the protostar 
decreases from $\sim 3R_{\ast}$ to $\sim 1R_{\ast},$ where $R_{\ast}$ is the radius of the 
star. In the standard CTTS model, the accreting matter is found in the magnetosphere for radii 
$\leq 3R_{\ast}.$ The presence of turbulence is assumed in the standard model for the 
accretion disk and we may assume that, due to the accreting matter, it is also present in 
the magnetosphere, where it can create large oscillations. In the region of the magnetosphere
which touches the magnetic field lines that thread the disk, these oscillations can generate 
AWs. The AWs could transport appreciable energy outward from the magnetosphere into the 
molecular cloud. 

\section{Acknowledgments}
The author would like to thank the Brazilian agencies FAPESP (Proc. No. 00/06770-2) 
and CNPq (Proc. No. 300414/82-0) as well as the project PRONEX (Proc. No. 41.96.0908.00) 
for partial support.


\begin{thebibliography}{60}
\bibitem{1} Balbus, S.A. and Hawley, J.F. (1991) ApJ, 376, 214.
\bibitem{2} Balbus, S.A. and Hawley, J.F. (1998) Rev. Mod. Phys., 70, 1.
\bibitem{3} Bavassano, B. (1990) Nuovo Cimento, 13, 79.
\bibitem{4} Edwards, S., Ray, T. and Mundt, R. (1993) in {\it Protostars and Planets III}
ed. E.H. Levy and J. Lunine Tucson: Univ. Arizona Press, 567.
\bibitem{5} Gammie, C.F. (1996) 457, 355.
\bibitem{6} Guenther, E.W., Lehmann, H., Emerson, J.P. and Staude, J. (1999) A\&A,
341, 768.
\bibitem{7} Hartmann, L., Hewett, R. and Calvet, N. (1994) ApJ, 426, 669.
\bibitem{8} Hawley, J.F. and Stone, J.M. (1998) ApJ, 501, 758.
\bibitem{9} Jatenco-Pereira, V., Opher, R., and Yamamato, L.C. (1994) ApJ, 432, 409.
\bibitem{10} Johns-Krull, C.M., Valenti, J.A., Hatzes, A.P. and Kanaan, A. (1999) ApJ,
   501, L41.
\bibitem{11} Lee, M.A. and Roberts, B. (1986) ApJ, 301, 430.
\bibitem{12} Martin, S.C. (1996) ApJ, 470, 537.
\bibitem{13} Muzerolle, J., Hartmann, L. and Calvet, N. (1998) ApJ, 492, 743.
\bibitem{14} Ouyed, R. and Pudritz, R.E. (1997) ApJ, 482, 712.
\bibitem{15} Parker, E.N. (1991) ApJ, 372, 719.
\bibitem{16} Stone, J.M., Hawley, J.F., Gammie, C.F. and Balbus, S.A. (1996) ApJ, 463, 656.
\bibitem{17} Tu, C.Y., Marsch, E. and Rosenbauer, H. (1990) Geophys. Res. Lett., 17, 283.
\bibitem{18} Tu, C.Y., Marsch, E. and Thieme, K.M. (1989) J. Geophys. Res., 94, 11739.  
\bibitem{19} Umebayashi, T. and Nakano, T. (1988) Prog. Theor. Phys. Suppl., 96, 151.
\bibitem{20} Vasconcelos, M.J., Jatenco-Pereira, V. and Opher, R. (2000) ApJ, 534, 967.
\bibitem{21} Vasconcelos, M.J., Jatenco-Pereira, V. and Opher, R. (2003) Astron. Ap., 384,
329.
\bibitem{22} Vitorino, B.F., Jatenco-Pereira, V. and Opher, R. (2002) Astron. Ap., 384,
329.
\bibitem{23} Vitorino, B.F., Jatenco-Pereira, V. and Opher, R. (2003) ApJ (in press).
\end{thebibliography}
\end{document}